\documentclass{osa-article}

\journal{osajournal}

\articletype{Research Article}
\usepackage{breqn}
\usepackage{semtrans}
\usepackage{float}

\begin{document}

\title{Mode expansion theory and application in step-index  multimode fibres for astronomical spectroscopy}

\author{E. Hernandez\authormark{1*}, M. M. Roth\authormark{1}, K. Petermann\authormark{2}, A. Kelz\authormark{1}, B. Moralejo\authormark{1} and K. Madhav\authormark{1}}

\address{\authormark{1}Leibniz Institut f\"ur Astrophysik Potsdam (AIP),
	An der Sternwarte 16, 14482 Potsdam, Germany\\
\authormark{2}Technische Universität Berlin, HFT4, Einsteinufer 25, 10587 Berlin, Germany}

\email{\authormark{*}ehernandez@aip.de} 



\begin{abstract}
In astronomical spectroscopy, optical fibres are abundantly used for multiplexing and decoupling the spectrograph from the telescope to provide stability in a controlled environment. However, fibres are less than perfect optical components and introduce complex effects that diminish the overall throughput, efficiency, and stability of the instrument. 
We present a novel numerical field propagation model that emulates the effects of modal noise, scrambling, and focal ratio degradation with a rigorous treatment of wave optics. We demonstrate that the simulation of the near- and far-field output of a fiber, injected into a ray-tracing model of the spectrograph, allows to assess performance at the detector level.
\end{abstract}


\section{Introduction}
\label{sec:intro}


In ground-based astronomy, optical fibres are still amongst the best and most flexible options to connect large and heavy instrumentation to the focal plane of a telescope. Instruments can be placed in the stable environment of a temperature and humidity controlled room. This facilitates the spectrograph optical alignment, reduces its sensitivity to motion and vibration from the telescope environment, and eases detector maintenance and operation. Fibres also offer input illumination stability, which is of fundamental importance for high-resolution spectroscopy. A few examples include HARPS, the extra solar planet hunter at the ESO La Silla 3.6m Telescope \cite{Pepe00}; the dual-channel visual and near-infrared echelle spectrograph CARMENES at the Calar Alto  3.5m Telescope \cite{Amado13}; the PEPSI echelle spectrograph at the Large Binocular Telescope (LBT) that operates at a resolving power of up to $R=250,000$ and allows for spectro-polarimetry  \cite{Strassmeier18}; and the ultra-stable echelle spectrograph ESPRESSO for the combined focus of the ESO Very Large Telescope (VLT), built with the goal to reach a radial-velocity precision of 10 $cm/s$ over a timespan of 10 years  \cite{Pepe20}.

Fibres enable the simultaneous recording of multiple targets in the field of view, a technique that is used for multi-object spectroscopy (MOS). State-of-the-art instruments feature from several hundred to thousands of fibres, e.g.  the 2dF-instrument at the Anglo-Australian Telescope (AAT), that was used for the 2dF galaxy redshift survey covering more than 100,000 galaxies \cite{Colless}; FLAMES at the ESO VLT that feeds two different spectrographs covering the whole visual spectral range and allowing up to 130 targets at a time  \cite{Pasquini02}; and FMOS at the Subaru Telescope, that uses an automated fibre positioner system to configure 400 fibres, covering the spectral range from 0.9$\mu m$ to 1.8$\mu m$ \cite{Kimura2010}. Arguably the most productive multi-object spectrograph to date is the fiber-fed Sloan Digital Sky Survey (SDSS) facility at the Apache Point Observatory \cite{York00}, now entering its 5th period after already 20 years of operation.

When fibres are configured as a bundle in the telescope focal plane they can sample spatially extended objects and can be re-arranged as a linear pseudo-slit configuration to feed a spectrograph. Through the use of such fiber bundle integral field units (IFUs), both the spatial and spectral information over the sampled field-of-view can be recorded simultaneously as a data cube, a technique that is known as integral field spectroscopy (IFS). Examples are VIRUS, built for the Hobby-Eberly Telescope Dark Energy Experiment (HETDEX) that uses 75 fibre bundle IFUs, each consisting of 448 optical fibers, with the entire system having around 34,000 individual fibres \cite{Hill08}; or PMAS installed at the Calar Alto 3.5m Telescope \cite{Roth05}. The retrofitted PPak-IFU has enabled the Calar Alto Legacy Integral Field Area Survey (CALIFA), yielding resolved spectroscopic information over  a 1 arcmin field-of-view in the wavelength range from $370nm$ to $700nm$ for 600 galaxies in the local Universe \cite{Sanchez12}. As this survey covers a statistically complete sample and reaches out to three effective radii of the observed galaxies, it has become a benchmark for similar surveys, that use multiplex capabilities thanks to deployable IFUs such as MAnGA \cite{Bundy2015} or SAMI \cite{Bryant2012}. 

Following these existing facilities, new fiber-fed instruments are being built or planned. As an example, the multi-object spectrograph 4MOST, currently under construction for the ESO Paranal Observatory VISTA telescope, will have more than 2400 science fibres aimed to conduct surveys that are expected to deliver over 7 million spectra \cite{deJong16}. Another next generation instruments under construction is MOONS, a multi-object spectrograph for the VLT \cite{MOONS}.

The next generation of extremely large telescopes will heavily rely on fiber-coupled spectrographs, e.g.\ ELT-MOS/MOSAIC for the ESO Extremely Large Telescope (ELT) \cite{Jagourel18}, GCLEF/\-GMACS with the fibre positioning system MANIFEST, designed for the Giant Magellan Telescope (GMT)  \cite{Colless2018}; and HIRES, the High Resolution Spectrograph for the ELT \cite{HIRES}. The list of fiber-fed spectrographs is by no means complete. However, it is clear that the technology of optical fibres remains of major importance for modern astrophysics. 

In this paper we investigate the end-to-end system performance of fibre-coupled spectrographs by presenting a wave-optical description of modes propagating along fibres, including light injection, and a description of the near-field and far-field distributions at the fibre output, both coupled into a ray-tracing model of the spectrograph's optical system. With proper data reduction software, the impact of fibre behaviour on the resulting spectra can be directly studied and quantified. Here, we focus on the fibre-related parts of the system. The spectrograph and data analysis part will be addressed in a follow-up paper. Section \ref{sec:fibprop} introduces fundamental properties of fibres that pose challenges to precision spectroscopy, Section \ref{sec:methods} describes the numerical methods, Section \ref{sec:simulations} presents results of si\-mulations and the experimental measurements, and a full end-to-end system simulation is illustrated in Section \ref{sec:application} followed by summary and conclusions in Section \ref{sec:summary}.

\section{Fundamental properties of fibres}
\label{sec:fibprop}

\subsection{Focal ratio degradation}
\label{subsec:FRD}

Despite their many benefits, fibres incorporate losses and unwanted effects that reduce the efficiency, throughput, and stability of the overall system. Non-conservation of \'{e}tendue causes focal ratio degradation (FRD) and results in a reduced throughput \cite{ramsey88}. It was first attributed to the broade\-ning of the angular distribution at the output of the fibre to the exchange of energy between the propagation modes caused by random irre\-gularities in the fibre \cite{gloge72}.  The term FRD, that is widely used amongst astronomers, but uncommon in other communities, was coined in the late 1970s when fibres were introduced in Astronomy \cite{Bacon}.  It was demonstrated that micro\-bendings, and poor fibre treatment, e.g. polishing or cleaving, can cause FRD \cite{ramsey88}.  The latter study concluded that the best focal ratio to feed a step-index multimode fibre is between f/3.0 and f/7.0.

Various studies found that macro- and micro-bending, mechanical stress, fibre alignment, perpendicularity between the fibre surface and fibre axis, polishing errors, and scattering effects within fibres have an influence on FRD, whereas no wavelength dependence could be observed 
 \cite{heacox92,Avila98, Schmoll03, Crause08}. 

\subsection{Modal noise}
\label{subsec:modalnoise}

The amplitude modulation of a signal in the fibre arising from modal interference caused by irregularities, misalignment, and limited coherence of source, is an undesirable effect in multimode fibres. According to \cite{Epworth79}, the conditions of modal noise are: 1) a sufficiently narrow source spectrum; 2) spatial or angular filtering; 3) time variation of either modal or spatial filtering. A rigorous evaluation of graded-index fibres to determine connector-introduced noise and distortions in optical links found that less coherence of the source diminishes both the fluctuations of coupling efficiency and the misalignment sensitivity of an optical link \cite{Petermann80}. 

An investigation of the distribution of power among the propagation modes showed that the coupling conditions dictate which modes will be excited, in contradiction to the preceding assumptions that all modes are excited equally \cite{Wood84}. By using the equations in \cite{Petermann80}, the signal-to-noise ratio (SNR) of an equally distributed modal power distribution (MPD) was computed and compared to realistic highly nonuniform MPD. The results demonstrated that a nonuniform MPD exhibits better SNR than an equally distributed MPD. 

An overview of the impact of modal noise on astronomical instrumentation is given in \cite{Lemke11}. In  \cite{Baudrand98}, modal noise was investigated as part of a conceptual design study for a high-resolution multi-aperture fibre-fed spectrograph. All the conditions for modal noise were given since: 1) the recorded spectral width was very narrow; 2) angular limitation occurred for the light injection and vignetting-obscuration took place in the spectrograph; 3) the fibre link moved with the telescope altering the modal and spatial filtering. When the fibres were perfectly still, the signal-to-noise ratio followed the expected photon statistics. However, when the fibre bundle was in stress or in motion, a significant degradation of the photometric performance was measured. An experimental study found that continuous agi\-tation of the fibre smooths out the modal pattern \cite{Baudrand01}. As explained in \cite{Lemke11}, agitation of the fibre induces temporary varying stress that causes phase scrambling among the propa\-gation modes.   

\subsection{Scrambling}
\label{subsec:scrambling}
Srambling describes the process in which the output distribution becomes independent from the input distribution of the fibre. It can be characterized as azimuthal or radial scrambling. According to \cite{heacox92}, it is bene\-ficial that the output field remains stable to isolate the spectrograph response from sources of systematic wavelength errors. With optimal modal scrambling, spectrographs can achieve the highest level of precision in radial velocity measurements \cite{Avila98, Hunter92, Barden93}, as needed e.g.\ for radial velocity exo-planet detections, and asteroseismology. In \cite{Avila10}, different fibre geometries were tested with different scrambling methods. Due to low efficiency,  beam homogenizers and field rotators were discarded as fibre scramblers. Non-circular fibres showed the best results when used in combination with mechanical fibre squeezers. The most known technique is fibre agitation \cite{Baudrand98, Baudrand01,Lemke10,Lemke11,Mccoy12 }. In all scrambling scenarios, energy among modes is being redistributed and can lead to an increase of FRD.

\subsection{Modelling fibres}
\label{subsec:simulations}
Simulating the effects of FRD, modal noise, and scrambling in multimode fibres for the optimization of instruments remains a critical issue for fibre-based astronomical systems. As new generation instruments are being designed for extremely large telescopes, whose light-collecting power and sensitivity reaches unprecedented values, the ever growing cost for focal plane instruments demands a good understanding of fibre properties, and performance issues imposed by their limitations. This situation has become more accute in the NIR wavelength regime, as the number of modes decrease with the inverse square of wavelength, but also with the growing size of ELT focus platforms, incurring long fiber lengths, hence attenuation and fibre motion, and other difficulties such as cryogenic system requirements. 

Light propagation in optical components can be simulated via ray tracing or wave analysis. Both methods can be used for simu\-lating the field propagation in optical fibres. Nevertheless, commer\-cially available software is not optimized to address the effects of FRD, modal noise or scrambling in multi\-mode fibres, such that user defined and modified simulations need to be applied for describing such effects. It is important to understand  that both the near-field and the far-field light distribution at the output of a fibre determine the shape of the image in the focal plane of the spectrograph, meaning that variations in these distributions will be reflected in changes of the line-spread-function (LSF) of the extracted spectra, hence the accuracy of spectra. The near-field, i.e.\ the illuminated fibre core, convolved with the abberations of the spectrograph optical system, is forming an image on the detector. The far-field that can be tracked in the pupil of the optical system, modulates the point-spread-function (PSF) of the system, hence aberrations, such that any changes will reflect in the image as well. An early attempt to model this fact was demonstrated in \cite{Schmoll03}. Varying aberra\-tions like astigmatism and coma near the edge of the field were clearly visible in the resulting LSF.  In \cite{Allington12}, FRD was simulated using a phase-tracking ray-tracing method implemented in MathCad. It was based on the model of \cite{gloge72} and represents a portion of the fibre with scattering centers to account for the deformation and irregularities a fibre may be subjected to. The scattering centers introduced perturbations in the ray path and distributed the rays into rays with larger or smaller propagation angles, i.e. into higher or lower order modes. The scrambling effects and modal noise (to an extent) could be analyzed with the model.

Since so many external parameters influence FRD, modal noise and scrambling, it is crucial to understand the intrinsic origin through the multimode propagation of light in optical fibres. In the simulations presented in this work, wave theory is used to describe and simulate these effects. The Eigenmode Expansion Method (EEM) is used as the basis to calculate the field propagation in multimode fibres. It creates the possibility to analyze the different modal effects on the near- and far-field of the fibre as a function of the input focal ratio, spatial and angular translation, focus variation, different wavelengths, and fibre geometries. A good agreement between simulation and experimental results based on analyzing the near- and far-field intensities, confirm that the numerical model is a good tool for characterizing fibres in a fibre-fed spectrograph systems.


\section{Numerical methods}
\label{sec:methods}

Field propagation in optical fibres can be simulated using the beam propagation method (BPM) or the Eigenmode expansion method (EEM).
In BPM, the computation time scales with the size of the device, and is not practical for modeling MM fibres used for astronomy. In EEM, the computational time scales with the number of modes $M$, and the resolution of the grid $\Delta x$ and $\Delta y$, where $x$ and $y$ are axes perpendicular to the axis of the fiber ($z$-axis). The EEM states that at any distance $z$ along the fibre, it is possible to use the propagation modes as a basis for linear expansion as  
\begin{equation}
\psi(x,y,z) = \sum_{m=0}^{M}c_{m}\psi_{m}(x,y)\exp\left( j\beta_{m}z\right),
\label{eq:eigenmode_expansion}
\end{equation}
where $\psi_{m}(x,y)$ is the electrical or magnetic field in a weakly guiding fibre, $\beta_{m}$ is the propagation constant, and $c_{m}$ is the modal amplitude of mode $m$. $|c_{m}|^{2}$ is the amount of power in each mode, and following conservation of power (normalized),
\begin{equation}
\sum^{M}_{m=0} |c_{m}|^{2} = 1.
\label{eq:power_conservation}
\end{equation}

Equation \ref{eq:eigenmode_expansion} shows that modes have a very simple harmonic z-dependence, which is the key to calculating the field propagation efficiently for long structures. Based on the EEM, a simulation environment has been developed that allows analyzing the modal power distributions (MPD) and near- and far-field distribution in simulated multimode step-index fibres of any length, geometry and with any kind of index difference. This allows for simulations with multiple degrees of freedom. 

\subsection{Calculating modes}
\label{sec:modes_calcuation}



To calculate the modes, the FemSIM module from the commercially available software RSoft was used. All other procedures and the mode expansion algorithms were coded in Python. For azimuthally invariant waveguides with small index difference $\Delta n$, such as a step-index fibre, the propagation modes are numerically calculated in the respective polarization states \cite{rsoft}.
We define the geometry of the fibre and parameters such as refractive index of the core ($n_{1}$) and  cladding ($n_{2}$), propagation wavelength $\lambda$, fibre diameter $2r$ and the number of modes $M\approx \frac{V^2}{2}$ , where $V=2\pi r NA/\lambda$ is the dimensionless frequency parameter.
Each mode is normalized to unity,
\begin{equation}
\iint |\psi_{m}(x,y)|^{2}\partial x \partial y = 1.
\label{eq:normalized_to_unity}
\end{equation} 
The Fraunhofer approximation is used to describe the far-field angular distribution. Through the Fourier transform,
\begin{equation}\label{eq:FT}
\mathcal{F}\left\{ \psi_{m}(x,y)\right\} = \Psi_{m}(\nu_{x},\nu_{y}),
\end{equation} 
the far-field information of each mode is produced. The angular distributions are obtained by u\-sing the approximations of $\theta_{x} \approx x/d \approx \lambda \nu_{x}$ and $\theta_{y} \approx y/d \approx \lambda \nu_{y}$ \cite{saleh}. For the following procedures, modes in angular coordinates are addressed as $\psi_{m}(\theta_{x}, \theta_{y})$, and modes in spatial coordinates are addressed as $\psi_{m}(x,y)$. 

\subsection{Calculating the input field}
\label{sec:Input_field}
The characteristics of the input field have a direct impact on the modes that are excited in a multimode fibre. Depending on the size, angle, and position, different modes will propagate. This will consequently produce a different MPD and affect directly the near- and far-field distributions. As a basic input field, a 2-D Gaussian distribution is used, 
\begin{equation}
\psi_{Gauss}(x,y) = A_{o}e^{ \big(-\frac{\sqrt{x^{2}+y^{2}}}{2\sigma^{2}_{o}}\big)}.
\label{eq:input_field}
\end{equation}
The factor $A_{o}$ is the amplitude and is set as a constant. The input field is normalized to unity as in Eq. \ref{eq:normalized_to_unity}. The FWHM = $2\sigma_{o}\sqrt{\ln2}$ is used as a measure of the input field size 
and the field radius of the distribution is $w_{o}=2\sigma_{o}.$
Spatial translation is applied through $\overline{x} = x -x_{o}$ and  $\overline{y} = y -y_{o}$ and the  incident angle is defined as $\psi_{\varphi} = \exp(j\varphi_{n}k_{o}n), n\in \{x,y\}$
where, $k_{o}= 2\pi / \lambda$ is the free space  wavenumber and $\varphi_n$ is the incident angle in the x- or y- axis. $n$ can be implemented in both axis simultaneously. The complete input function can be described as,  
\begin{align}
\psi_{in}(\overline{x},\overline{y}) &= \psi_{Gauss}(\overline{x},\overline{y})\cdot\psi_{\varphi}, \nonumber \\
\nonumber \\
 &=  A_{o} e^{\big(-\frac{\sqrt{\overline{x}^{2}+\overline{y}^{2}}}{2\sigma^{2}_{o}}\big)}e^{j\varphi_{n}k_{o}n}, n\in \{x,y\}.
\label{eq:complete_input}
\end{align}

\subsection{Calculating the modal amplitudes}
\label{sec:modal_amplitudes}

As described in Eq. \ref{eq:eigenmode_expansion}, the modal amplitudes are represented with the parameter $c_{m}$. For $z=0$, Eq. \ref{eq:eigenmode_expansion} equals
\begin{equation}
\psi(x,y,z=0) = \sum_{m=0}^{M} c_{m}\psi_{m}(x,y)e^{j\beta_{m}0}, 
\label{eq:input_z_equals_zero}
\end{equation}
which basically represents the input field, 
\begin{equation}
\psi_{in}(\overline{x},\overline{y}) = \sum_{m=0}^{M}c_{m}\psi_{m}(x,y).
\label{eq:input_filed_z_0}
\end{equation}
Since the propagation modes are orthogonal to each other, 
\begin{equation}
\langle \Psi_{m},\Psi_{n}^{*} \rangle = \iint \psi_{m} \psi_{n}^{*} \exp \left(  j(\beta_{m} -\beta_{n} )z \right) \partial x \partial y = 0, \text{ for } m \ne n,
\label{eq:orthonormal_modes} 
\end{equation}
we can use Eq. \ref{eq:input_filed_z_0} and Eq.  \ref{eq:orthonormal_modes} to calculate the modal amplitudes through an overlap integral using the input field and the transverse modal fields,
\begin{equation}
c_{m} = \iint \psi_{in}(\overline{x},\overline{y}) \psi_{m}^{*}(x,y) \partial x \partial y.
\label{eq:overlap_int}
\end{equation}

\begin{figure}[htb]
	\centering
	\includegraphics[scale=0.46]{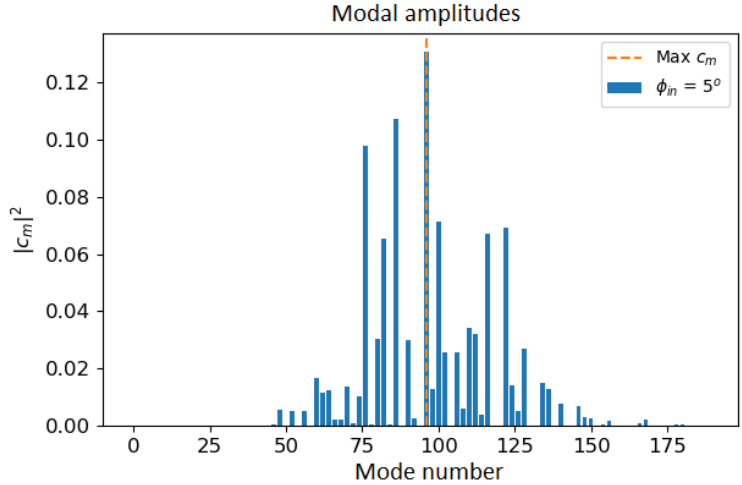}
	\caption{Modal power distribution of circular fiber with $M=180$: $r=50\mu m$, and  $\text{NA} \approx 0.121$ at $\lambda=1\mu m$.}
	\label{fig:modal_amp_example}
\end{figure}

The total amount of power being injected into the fibre is divided among the active modes. The modal amplitude array,  
\begin{equation}
c = [c_{o}, c_{1}, c_{2},\dots, c_{M}] ,
\label{eq:modal_amplitude_array}
\end{equation}
contains the MPD \cite{Corbett07} and can be used to illustrate the active modes of a system. As an example, Fig. \ref{fig:modal_amp_example} illustrates the MPD of a simulated fibre with an NA = 0.121, $r = 25\mu m$ and $\lambda=$ 1.0$\mu m$, giving $V=18.96$ and $M=180$. The simulated input field was a Gaussian with FWHM = $25\mu m$ and $\phi_{in}=5^{o}$, where $\phi_{in}$ represents $\varphi_n$ from Eq. \ref{eq:complete_input}. The modal amplitude array is normalized as described in Eq. \ref{eq:power_conservation}.

The active propagation modes in the system are a function of the characteristics of the input field $\psi_{in}(\overline{x},\overline{y})$ and not all modes are excited equally \cite{Wood84}. For the mode which carriers the max power, $c_{max}$, the calculated output angle is $\phi_{cmax,out} = 5^{o}$, which is same as the input angle. Figure \ref{fig:modal_amp_example} shows that modes with higher order than $c_{max}$ are also excited. Although no perturbations of the system have been taken into account, this indicates the possibility that the output beam will have a higher divergence than the input, i.e. FRD. Even though a complete assessment on the divergence of the output beam can be done only after a complete simulation, the knowledge of the active modes in a system already gives insights into the expected divergence.

The NA is dependent on $n_1$ and $n_2$ of the fibre (NA = $\sqrt{n_1^2-n_2^2}$). Since each mode has a specific exit angle \cite{gloge72} and possesses an effective refractive index $n_{e\!f\!\!f,m}$, each mode can be given an effective numerical aperture as in NA$_{e\!f\!\!f}(m)$. In the following example, the value $\text{NA}_{e\!f\!\!f}(m)$ will be used to assess the divergence of the output angle depending on the active propagation modes. 

Provided the spatial distribution of the input field fills the core of the fibre, it is possible to use the underfilled NA of the fibre to estimate the number of excited modes (2018, priv. comm. Prof. Tim Birks and Dionne Haynes). This can be applied to the $\text{NA}_{e\!f\!\!f}(m)$ if expressed as a function of the number of propagation modes using $V$ and $M$,
\begin{equation}
\text{NA}_{e\!f\!\!f}(m) = \frac{\lambda \sqrt{2m}}{2\pi r}
\label{eq:NA_function_modes}
\end{equation}
Furthermore, using the effective refractive index of each mode directly, $\text{NA}_{e\!f\!\!f}(m)$ can be calculated via
\begin{equation}
\text{NA}_{e\!f\!\!f}(n_{e\!f\!\!f\!,m}) = \sqrt{n^{2}_{1} - n^{2}_{e\!f\!\!f\!,m}}
\label{eq:effective_NA_refractive_index}
\end{equation} 
Using Eq. \ref{eq:FT} we obtain the angular distribution in the far-field $\Psi_{m}(\theta_{x},\theta_{y})$, thus allowing for the approximation of the output NA as a function of $\theta_{x},\theta_{y}$. This method is analogous to \cite{snyder} which showed that for $M \gg 1$, each mode can be associated with a unique characteristic ray angle $\theta_{in,out}$ \cite{Corbett07}. 

\begin{figure}[htb]
	\centering
	\includegraphics[scale=0.26]{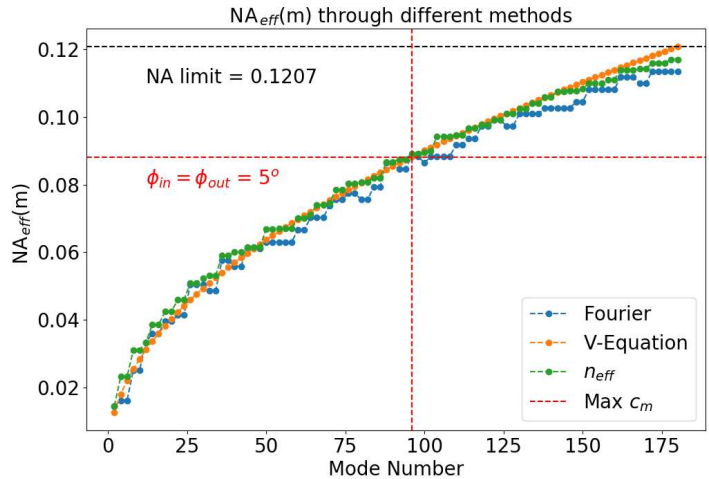}
	\caption{Calculation of $\text{NA}_{e\!f\!\!f}(m)$ as a function of the propagation modes using three different methods: Fraunhofer approximation, V-Equation and $n_{e\!f\!\!f\!,m}$. $\text{NA}_{e\!f\!\!f}(m)$ can be used as a measure for the output beam divergence. The mode carrying the maximum power in Fig. \ref{fig:modal_amp_example} with the respective output angle $\phi_{out}$ is marked.}
	\label{fig:approximation_different_methods}
\end{figure}

When the modes are analyzed in the far-field, the intensity does not have a sharp boundary that marks a specific output  angle. After Fourier transform, the angular intensity field was averaged to a 1D intensity distribution. The maximum of the averaged 1D distribution was used to mark the output angle of the field. If the mode is from the $LP_{0x}$ family, then the outermost ring is evaluated. Figure \ref{fig:approximation_different_methods} shows the results of the three different methods for $M=$180 modes presented in Fig. \ref{fig:modal_amp_example} with the simulated fibre. The modal number carrying the most power with the corresponding $\phi_{cmax,out}$ is marked.

As can be seen in Fig. \ref{fig:approximation_different_methods}, the three different methods correlate with the prediction that increasing mode numbers will increment $\text{NA}_{e\!f\!\!f}(m)$ and produce more divergence at the output end of the fibre. It is analog to the description in \cite{Lemke11} that the number of propagation modes rises with an increment in $\phi_{in}$. Since the number of propagation modes is a function of the input angle, $M(\phi_{in})$, and the divergence of the output beam depends on the number of propagation modes, Fig. \ref{fig:approximation_different_methods} shows how input angular fluctuations could translate directly into far-field angular fluctuations, a situation that can result in photometric errors in a fibre-fed spectrograph system.  

\subsection{Generating the output fields}
\label{sec:NF_FF_section}
After the modal amplitudes are calculated, the summation of all propagation modes at the distance $z=L$ provides the resulting field. The superposition between the modes can be coherent, partially coherent or incoherent. For the time being only the cases of complete coherence and complete incoherence are taken into account. 

When the calculation of the field is completely deterministic, in other words coherent, the near-field output intensity is obtained from the absolute square of Eq. \ref{eq:eigenmode_expansion} as
\begin{equation}
I_{out}(x,y,L) = \left|\Psi_{out}(x,y,L) \right|^{2} = \left| \sum_{m=0}^{M}c_{m}\psi_{m}(x,y)\exp(j\beta_{m}L)\right|^{2}. 
\label{eq:intensity_nearfield_coherent}
\end{equation}

During scrambling, the phases of the modes are completely randomized, averaging the phase factor between modes to zero. This means that the intensity of the propagation modes can be added to obtain the near-field output intensity and thus making the field incoherent:
\begin{equation}
I_{out}=\sum_{m=0}^{M}I_{m} = |c_{o}\psi_{o}|^{2}+|c_{1}\psi_{1}|^{2}+...+|c_{M}\psi_{M}|^{2}.
\label{eq:nearfield_intensity_incoherent}
\end{equation}

The superposition of modes in the far-field follows the same rules as for the near-field. As described in Sec. \ref{sec:modes_calcuation}, the Fraunhofer approximation is used for calculating the angular far-field distribution. Through the Fourier transform of the near-field, $\Psi(\nu_{x},\nu_{y}) = \mathcal{F}\{\Psi_{out}(x,y)\}$, and applying $\theta_{x,y} \approx \lambda\nu_{x,y}$, the angular distribution is obtained. The intensity of the angular distribution is equivalent to 
\begin{equation}
I(\theta_{x},\theta_{y}) = \left| \Psi(\theta_{x},\theta_{y}) \right|^{2}.
\label{eq:farfield_intensity_coherent}
\end{equation}  

For computing the coherent case a 2D-FFT can be applied to the near-field distribution. Another way is by transforming each mode individually and applying the coherent superposition in the angular plane. Since the transform of a mode can be expressed as 
\begin{equation}
\mathcal{F}\left\{ c_{m} \psi_{m} e^{j\beta_{m}z} \right\} = c_{m}e^{j\beta_{m}z}\iint\psi_{m}(x,y) e^{-j2\pi (\nu_{x}x+\nu_{y}y)}\partial x \partial y,
\label{eq:fourier_transform_individual_mode}
\end{equation} 
a single mode in the far-field angular plane can be expressed as $c_{m}e^{j\beta_{m}z}\mathcal{F}\{ \psi_{m}\}$. Using the linearity superposition property of the Fourier Transform $\mathcal{F}\{ax(t)+by(t)\} = a\mathcal{F}\{x(t)\} +b\mathcal{F}\{y(t)\}$, we get the coherent far-field as 
\begin{align}
\mathcal{F} \left\{ \Psi_{out} \right\} &= \mathcal{F}\left\{ \sum_{m=0}^{M} c_{m} \psi_{m} e^{j\beta_{m}z} \right\} \nonumber \\
\nonumber \\
 &=  c_{0}e^{j\beta_{0}z}\mathcal{F}\left\{ \psi_{0} \right\}  + ... + c_{M}e^{j\beta_{M}z}\mathcal{F}\left\{ \psi_{M} \right\}.
\label{eq:linearity_EEM_farfield}
\end{align}
$\Psi_{out}(x,y)$ and $\psi_{m}(x,y)$ were written without the argument $(x,y)$ for ease of  displaying the equations. Equation \ref{eq:linearity_EEM_farfield} is used for si\-mulating coherent propagation in the angular plane because it facilitates the computation of multiple simulations. 
Since the modal amplitudes are already calculated for the near-field, no additional steps are required, thus simplifying computations for each far-field. 
Analog to Eq. \ref{eq:nearfield_intensity_incoherent}, the incoherent far-field intensity can be calculated with 
\begin{align}
\left| \mathcal{F}\left\{ \Psi_{out}  \right\}\right|^{2} &= \sum_{m}^{M} |\mathcal{F}\left\{ c_{m}\psi_{m}  \right\}|^{2} \nonumber \\
\nonumber \\
&= \left| \mathcal{F}\left\{ c_{0} \psi_{0}  \right\} \right|^{2}  + \left| \mathcal{F}\left\{ c_{1} \psi_{1}  \right\} \right|^{2} +... + \left|\mathcal{F}\left\{ c_{M} \psi_{M}  \right\} \right|^{2}.
\label{eq:incoherent_far_field} 
\end{align} 

\subsection{Mode coupling}
\label{sec:mode_coupling_section}
If the input field and the modal distributions of a system are known, the MPD of a system can be obtained. In a real fibre, due to perturbations in the refractive index, the MPD changes through the propagation length, i.e,  MPD(z=0) $\neq$ MPD(z=L). The modal amplitudes interchange power (mode coupling) as they propagate along the waveguide, making the modal amplitude array  a function of the propagation distance,  $c(z)$. The modal amplitudes are no longer constant and Eq. \ref{eq:eigenmode_expansion} can be rewritten as

\begin{equation}
\Psi_{out}(z) = \sum_{m=0}^{M}c_{m}(z)\psi_{m}e^{j\beta_{m}z},
\label{eq:mode_coupling}
\end{equation}	
To solve for the modal amplitudes, it is helpful to use the simplified ordinary differential equation (ODE) \cite{Huang94}
\begin{equation}
\frac{\partial c_{\nu}}{\partial z} = \sum_{\nu} j \kappa_{\nu \mu} c_{\nu} e^{j(\beta_{\mu} - \beta{\nu})z},
\label{eq:ODE_mode_coupling}
\end{equation}
and describing the coupling coefficients as
\begin{equation}
\kappa_{\nu \mu} = \omega \iint \psi_{\nu}^{*}\Delta n^{2} \psi_{\mu} \partial x \partial y.
\label{eq:coupling_coefficinets} 
\end{equation}
The factor $\Delta n^{2}$ is a function that describes the perturbations in the refractive index. For a numerical calculation of $\kappa_{\nu \mu}$, an appropriate $\Delta n^{2}$ function needs to be found. This is fairly difficult since it would be based on perturbation theory and depends on many input parameters that go beyond the scope of this work. 
Hence, a phenomenological approach was used. 

In a regular multimode fibre, mode coupling occurs mostly between neighboring modes and the coupling strength decreases  with the mode spacing. This has been tested experimentally and described theoretically in \cite{gloge72}, where  the modes of a multimode fibre were analyzed as a continuum and mode coupling was described as a diffusion process. In \cite{Allington12} it was stated that the effects described by Gloge postulate the existence of scattering sites distributed within the fibre, which scatter light into higher- and lower-order modes. As explained in  \cite{Haynes11}, the scattering points are concentrated in the termination points of the fibre. Mechanical and thermal stress caused by the finishing processes (e.g. polishing, grinding and/or cleaving) is believed to introduce micro-ruptures into the core and cladding causing abrupt refractive index changes $\Delta n$ that are the source of mode coupling. The factor $\Delta n$ can be seen in Eq. \ref{eq:coupling_coefficinets} and is required to define the coupling coefficient between two propagation modes. 

To reproduce a similar procedure, the modal amplitude array $ c = \left\{ c_{0},c_{1},...,c_{M} \right\} $  is convolved with a 1-D Gaussian function $ \Gamma_{\sigma} = e^{(-x^{2}/2\sigma^{2})} $. The number of elements in $c$ is the same as for $\Gamma_{\sigma}$.  Having a FWHM = $2\cdot\sqrt{2\ln 2}\sigma$, the factor $\sigma$ in  $\Gamma_{\sigma}$ indicates the weight of mode coupling used in the simulation. The values for $\Gamma_{\sigma}$ are normalized to 1. It results in $c_{\sigma} = c\circledast \Gamma_{\sigma}$, with $c_{\sigma}$ still holding the power conservation with $ \sum_{m=0}^{M}|c_{\sigma}|^{2} = 1$. 

\begin{figure}[htb]
	\centering
	\includegraphics[scale=0.4]{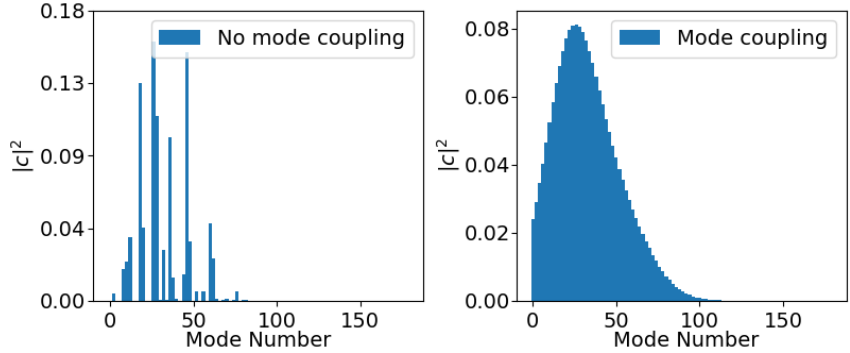}
	\caption{Example of the modal power distribution before and after applying mode coupling to the simulations. A discrete power distribution is observed without mode coupling, and a continuum after mode coupling.}
	\label{fig:mode_coupling}
\end{figure}

The calibration of $\sigma$, and thus the amount of FRD, is done by using actual measurements and finding the appropriate va\-lues for the simulations. Using the same fibre described above, Fig. \ref{fig:mode_coupling} shows the modal power distribution before and after applying mode coupling. The simulated input field was a Gaussian with a FWHM = 25$\mu m$ and $\phi_{in} = 3^{o}$. A discrete power distribution is observed without mode coupling. A continuum in the power distribution if observed after applying mode coupling.

\section{Results}
\label{sec:simulations}

This section illustrates examples of simulated output fields. First, a comparison between simulated fields and measured output fields is presented. Afterwards, a more quantitative measure is used which is based on the power spectral density of the speckle patterns produced by the near-fields. And last, an ana\-lysis of the angular divergence and coupling efficiency is presented. 

Two fibres with a high number of propagation modes were simulated:  circular and octagonal cross-section. All the other parameters are identical. The simulated fibres have an NA = 0.22, $r=25\mu m$, $L=2$m, $\lambda=633 nm$, giving $V=$ 54.59 and $M=1490$.

The measured fibres used for comparison had close to identical parameters for the NA, $r$ and $L$. A HeNe laser at $633nm$  was used for the experiments, producing the same $V$ and $M$ as the simulated fibres. The core/clad materials were silica and doped silica, allowing for a low loss bandwidth between $ 200nm - 2100nm$.  The circular fibre was manufactured by Polymicro Technologies, Molex\textsuperscript{\textregistered} and had the batch identification FBP050070085. The octagonal fibre was manufactured by Ceram Optec\textsuperscript{\textregistered} and had the batch identification OCTWF50/94P. More information about the fibres and the experimental measurements can be found in the supplementary document.

\subsection{Comparison between simulations and measurements}

\subsubsection{Near-field}

\begin{figure}[htb]
	\centering
	\includegraphics[scale=0.38]{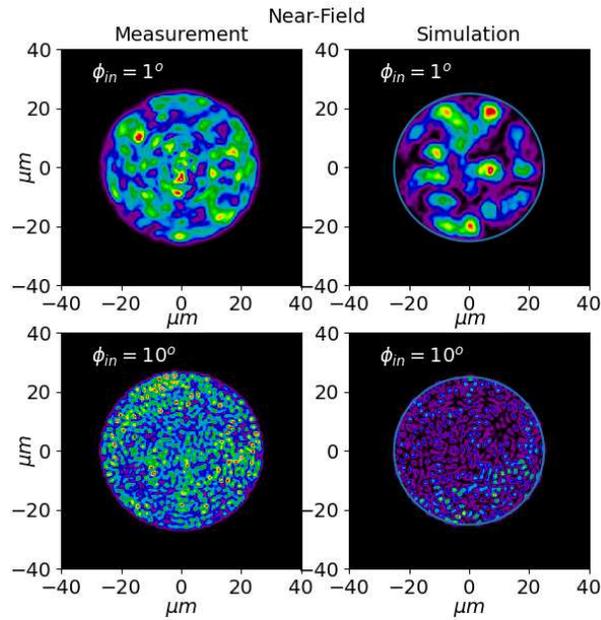}
	\caption{Comparison of the near-field intensity between the simulated circular fibre and measured circular fibre for two different incident angles of $\phi_{in} = 1^{o}$ and $\phi_{in} = 10^{o}$.}
	\label{fig:near_field_circular_fibre}
\end{figure}

\begin{figure}[H]
	\centering
	\includegraphics[scale=0.38]{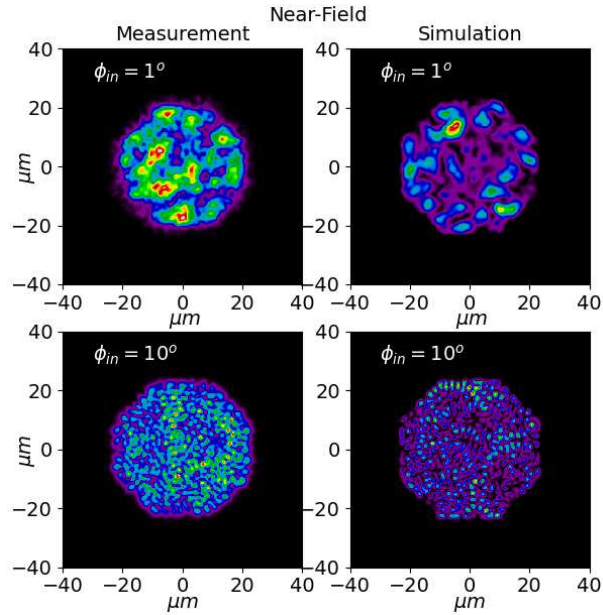}
	\caption{Comparison of the near-field intensity between the simulated octagonal fibre and measured octagonal fibre for two different incident angles of $\phi_{in} = 1^{o}$ and $\phi_{in} = 10^{o}$.}
	\label{fig:near_field_oct_fibre}
\end{figure}

Figure \ref{fig:near_field_circular_fibre} illustrates the near-field of the circular fibre for $\phi_{in} \in 1^{o}, 10^{o}$. In this case, coherent propagation is simulated, meaning Eq. \ref{eq:intensity_nearfield_coherent} is applied. The simplified mode coupling procedure explained in Sec. \ref{sec:mode_coupling_section} is used for the simulations and set to $\sigma = 0.07$. 
The value of $\sigma$ is further analyzed in Sec. \ref{sec:power_spectral_density_section} and Sec. \ref{sec:angular_divergence_simulations}. Figure \ref{fig:near_field_oct_fibre} illustrates the same procedure for the octagonal fibre. For a better visibility of the speckles, the spectral color map was used.  

For both fibres the number of speckles in the near-field increases with the incident angle $\phi_{in}$. The number of modes that propagate in the system can be approximated by the number of speckles that can be counted in the near-field distribution \cite{Lemke11}. But the power is not distributed equally among propa\-gation modes \cite{Wood84}. In fact, the coupling conditions dictate which modes will be excited. In this case, the steeper incident angles, e.g. $\phi_{in} = 10^{o}$, excite higher order modes. The intensity fluctuations that result from the interference of higher order modes have high spatial frequencies, thus inducing a higher number of speckles when $\phi_{in}$ increments. This will be quantified in Sec. \ref{sec:power_spectral_density_section} with the power spectral density analysis. 

\begin{figure}[htb]
	\centering
	\includegraphics[scale=0.4]{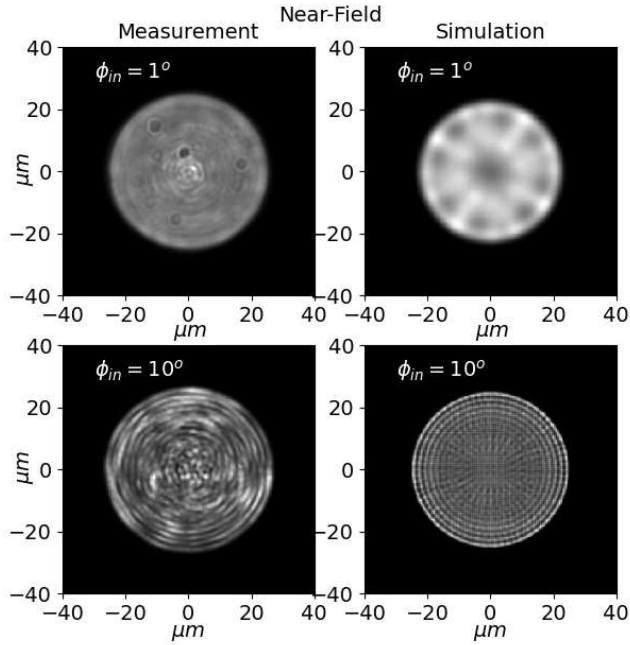}
	\caption{Comparison of the near-field intensity between the simulated circular fibre and measured circular fibre for two different incident angles of $\phi_{in} = 1^{o}$ and $\phi_{in} = 10^{o}$. Incoherent propagation is applied in this example.}
	\label{fig:near_field_circular_fibre_incoherentt}
\end{figure}

\begin{figure}[htb]
	\centering
	\includegraphics[scale=0.4]{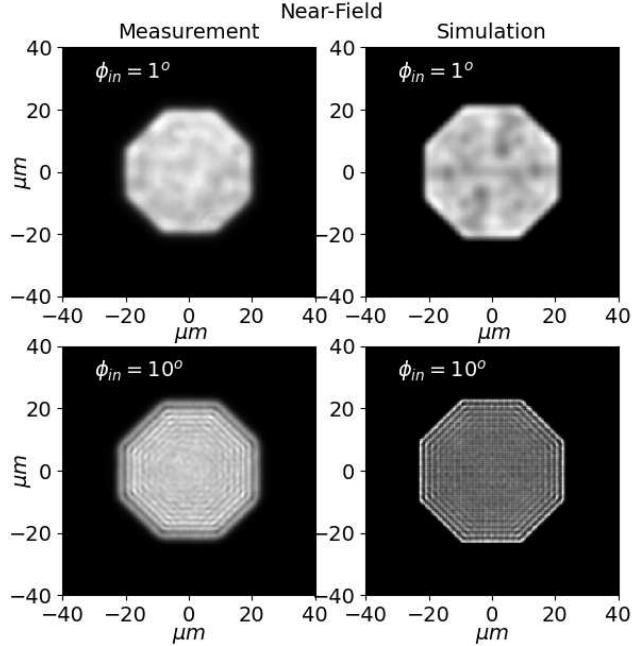}
	\caption{Comparison of the near-field intensity between the simulated octagonal fibre and measured octagonal fibre for two different incident angles of $\phi_{in} = 1^{o}$ and $\phi_{in} = 10^{o}$. Incoherent propagation is applied in this example.}
	\label{fig:near_field_oct_fibre_incoherent}
\end{figure}

Modal interference is the main cause of modal noise and introduces difficulties for both near- and far-field analysis. To smooth out the speckle pattern, modal scrambling can be induced. Perfect modal scrambling occurs when the phases of all active modes are completely randomized. As indicated in Sec. \ref{sec:intro}, one way of applying induced scrambling to an optical fibre is by agitating the fibre. Numerically, this means applying the case of incoherent propagation, i.e, using Eq. \ref{eq:nearfield_intensity_incoherent} for simulating the near-field. The next examples show the near-field for the case of induced scrambling. Once more, two different incident angles are illustrated, $\phi_{in} = 1^{o}$ and $\phi_{in} = 10^{o}$, and the simplified mode coupling procedure is used for the simulations. These results are presented in Fig. \ref{fig:near_field_circular_fibre_incoherentt} for the circular fibre and in Fig. \ref{fig:near_field_oct_fibre_incoherent} for the octagonal fibre. For a better visibility of the periodic intensity fluctuations by the larger input angles, we use a gray scale.

Although the speckles vanished in both cases, intensity patterns of different spatial frequencies appeared as a function of the input angle $\phi_{in}$. The simulations can reproduce the measurements of the octagonal fibre very accurately. The low spatial frequency fluctuations for the case of $\phi_{in} = 1^{o}$ are visible for both  the measurement and  the simulation. The fluctuations are caused by the fields of the excited modes being summed incoherently. The same principle applies for the case of $\phi_{in} = 10^{o}$. The visible structures are caused by the fields of the higher-order modes being summed incoherently (scrambled).

In the case of the circular fibre, the simulations are slightly differed from the measurements. Low spatial frequency fluctuations can be observed, but also higher frequency fluctuations are visi\-ble in the measurement. This may be due to degenerate mode groups which exist in circular fibres and which - in the strict sense - are not allowed to be superimposed incoherently. Fluctuations resembling coherent interference patterns are visible. This can described with the case of partial coherence. The contrast of an interference pattern is given by the visibility factor 
\begin{equation}
\nu = \frac{I_{max} - I_{min}}{I_{max} + I_{min}}. 
\label{eq:visibility}
\end{equation} 
For a multimode fibre, the visibility factor can be determined by the fibre-mode density and auto-correlation function of the source \cite{Kanada83}. Between the propagation modes, the degree of coherence $\gamma_{nm}$ decreases due to modal dispersion, with $n$ and $m$ being two different modes. In other words, $\gamma_{nm}$ is analogous to the visi\-bi\-lity factor $\nu$ between the individual modes. 

The following points for describing the partial coherence relation between the individual modes are still under study. To solve for $\gamma_{nm}$, the difference in group velocity between the modes have to be calculated. In the specific case of the step-index fibre, the difference in group velocity equals the phase delay difference with an opposite sign, allowing to calculate   
\begin{equation}
\Delta L = \frac{\lambda \Delta \beta_{nm}}{2\pi}z_{pr\!o\!p}.
\label{eq:interferometricpathmodes}
\end{equation}
The degree of coherence $\gamma_{nm} (\Delta L)$ is proportional to the Fourier transform of the source spectral density \cite{Phd_Bruening}
\begin{equation}
\gamma_{nm}(\Delta L) = \left| \frac{\int_{0}^{\infty}\partial \nu S(\nu) \exp(-2\pi j\nu \Delta L/c)}{\int_{0}^{\infty} \partial \nu S(\nu)}\right|.
\label{eq:degree_of_coherence}
\end{equation}
For the implementation of partial coherence in the simulations, the degree of coherence $\gamma_{nm} (\Delta L)$ needs to be acquired for all the possible modal combinations.We are planning to include this feature in a future upgrade of the code.
 
\subsubsection{Far-field}

The far-field simulations are realized to resemble the far-field of the parallel laser beam method, which consists of injecting collimated light into the fibre at an input angle $\phi_{in}$. The far-field response produces a ring of finite width $\Delta \theta$ that spreads under the angle $\phi_{out} = \phi_{in}$. This method is typically used for analyzing the FRD in a fibre by measuring the  finite width $\Delta \theta$ of the ring as a rapid diagnostic tool. Each input angle $\phi_{in}$ can be related to a specific set of propagation modes \cite{gloge72}, meaning that the far-field angular distribution directly represents the modal power distribution. If perturbations occur to a specific modal power distribution, $\Delta \theta$ experiences broadening equivalent to an increase in FRD in the system. This allows the use of a simplified mode coupling procedure to emulate disturbances in the system by giving power to modes that are originally not excited. Increasing the mode coupling parameter represents more disturbances and results in the broadening of $\Delta \theta$.  

\begin{figure}[htb]
	\centering
	\includegraphics[scale=0.52]{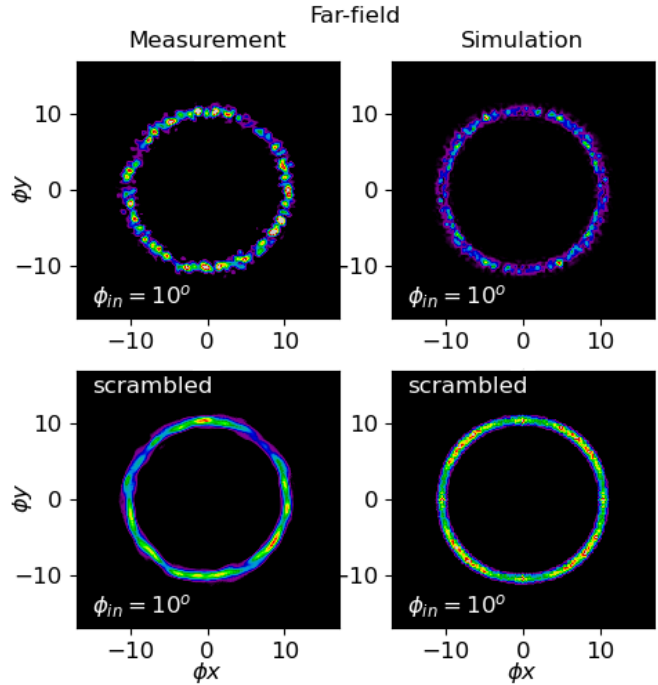}
	\caption{Comparison of the far-field intensity distribution between the simulated circular fibre and measured circular fibre for $\phi_{in} = 1^{o}$ and $\phi_{in} = 10^{o}$. Incoherent propagation is applied in the example marked as scrambled. All the axes are in degrees for the angular coordinates $\phi_{x}$ and $\phi_{y}$.}
	\label{fig:FF_circular_fibre}
\end{figure}

\begin{figure}[htb]
	\centering
	\includegraphics[scale=0.52]{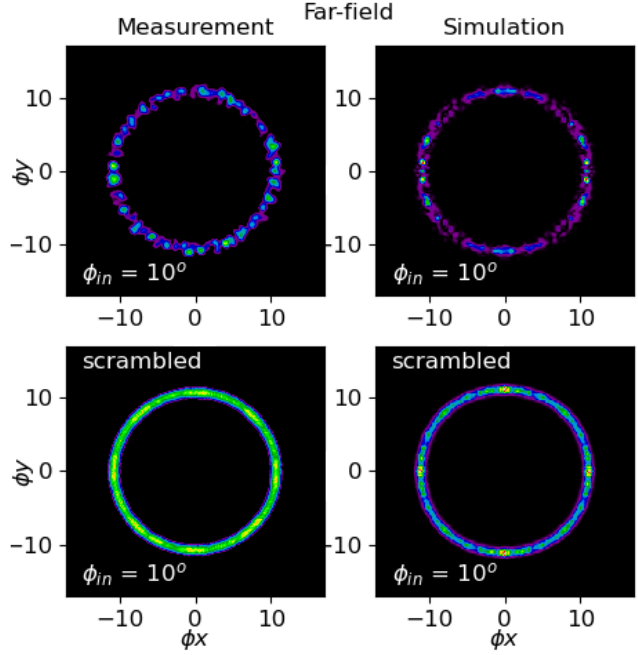}
	\caption{Comparison of the far-field intensity distribution between the simulated octagonal fibre and measured octagonal fibre for the incident angle of $\phi_{in} = 10^{o}$. Incoherent propagation is applied in the example marked as scrambled. All the axes are in degrees for the angular coordinates $\phi_{x}$ and $\phi_{y}$.}
	\label{fig:FF_octagonal_fibre}
\end{figure}

Figure \ref{fig:FF_circular_fibre} and Figure \ref{fig:FF_octagonal_fibre} show simulations of the far-field when using the parallel laser beam technique with the circular and octogonal fibre, respectively, for $\phi_{in} = 10^{o}$. Both the non-scrambled (coherent, Eq. \ref{eq:linearity_EEM_farfield}) and the scrambled (incoherent, Eq. \ref{eq:incoherent_far_field}) cases are shown.  The simplified mode coupling procedure is used for the simulations. The axes represent the angular coordinates $\phi_{x}$ and $\phi_{y}$ in degrees.
 
\subsection{Power spectral density analysis}
\label{sec:power_spectral_density_section}
\begin{figure}[htb]
	\centering
	\includegraphics[scale=0.28]{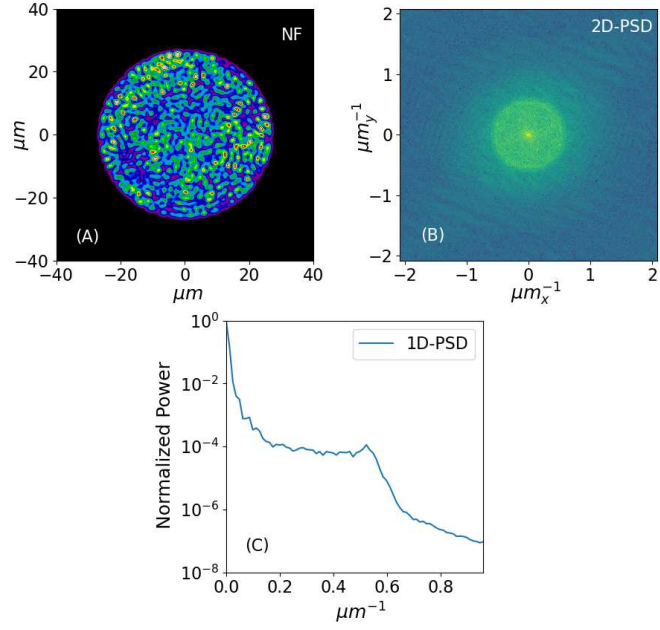}
	\caption{Example of the PSD of a given intensity field. (A) The NF represents a measured near-field distribution. (B) The 2D-PSD represents the autocorrelation of the field in the Fourier plane. (C) The 2D-PSD is the rotationally averaged and normalized.}
	\label{fig:PSD_example}
\end{figure}

As can be observed in Fig. \ref{fig:near_field_circular_fibre} and in Fig. \ref{fig:near_field_oct_fibre}, with increasing $\phi_{in}$, the number of speckles in the near-field intensity distribution increase. This occurs due to the interference of high-order modes that produces higher spatial fluctuations. To prove this statement it is possible to use the power spectral density (PSD) analysis. The PSD of a field disaggregates  the intensity into individual contributions from the different spatial frequencies. Mathematically, it is the Fourier transform of the autocorrelation function of the intensity field, which contains power across a range of wavevectors \cite{Jacobs17}. An increase in spatial frequency indicates existence of smaller elements in the measured intensity field. It applies to a higher number of speckles and a greater rate of intensity fluctuations inside the field. 
We calculate the fast Fourier transform (FFT) of the field and compute the square of the absolute value to get the PSD, i.e, 


\begin{equation}
PSD = |\mathcal{F}\{I(x,)\}|^{2} = |I(\nu_{x},\nu_{y})|^{2}.
\label{eq:PSD}
\end{equation} 
where, $I(x,y)$ is the intensity of a given field $\Psi(x,y)$.
The spatial frequency image is afterwards rotationally averaged to produce the normalized power over spatial frequency 1D-plot. Figure \ref{fig:PSD_example} illustrates first an intensity field, then the 2D autocorrelation, and lastly the $\mathcal{I}(\nu)$ 1D-PSD of the field. 

\begin{figure}[H]
	\centering
	\includegraphics[scale=0.27]{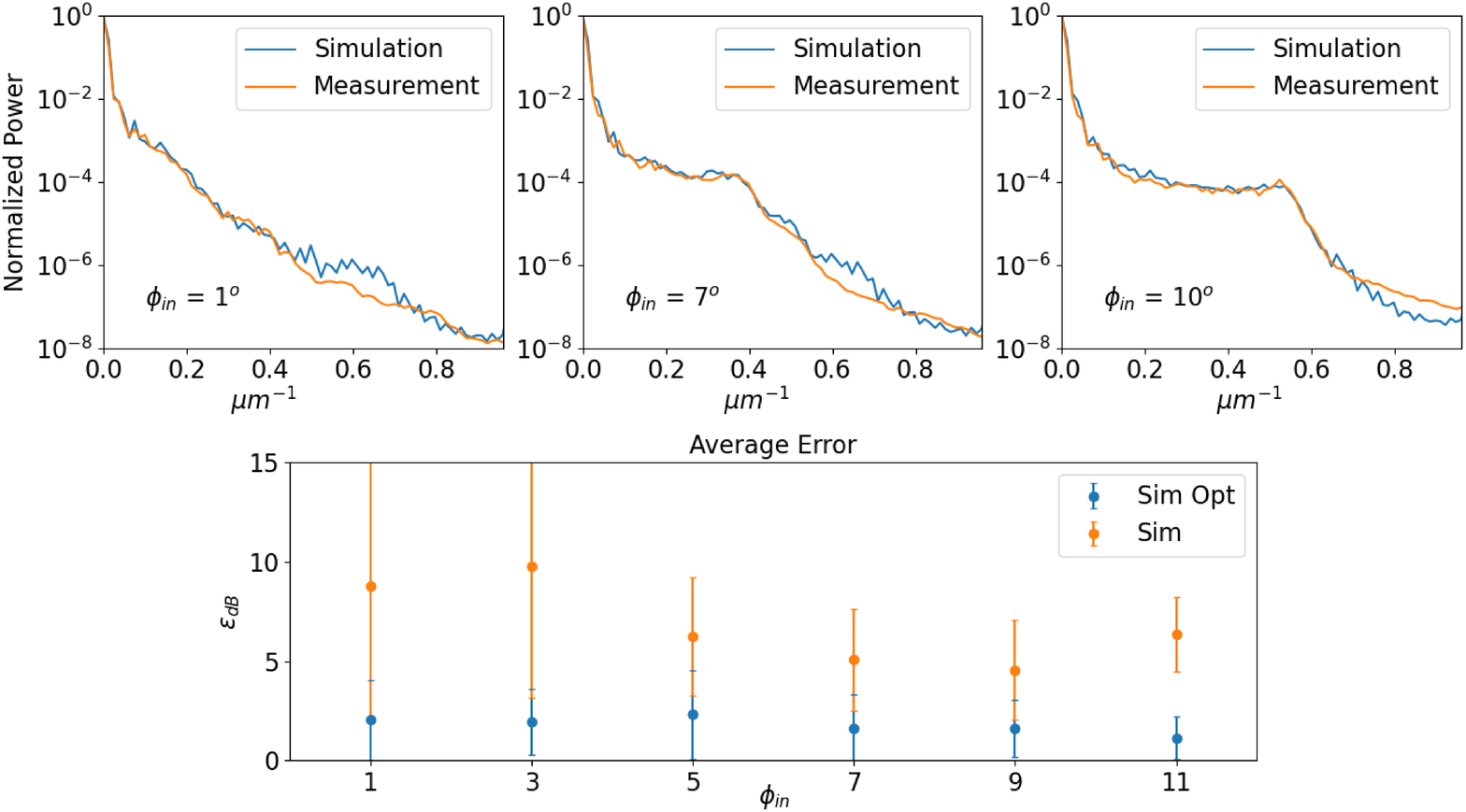}
	\caption{\textbf{(Top)} Comparison of power spectral density for measurements and simulations with incident angles of $\phi_{in} \in \{1^{o},7^{o},10^{o} \}$. The x-axis represents the spatial frequency and the y-axis is the normalized power in the logarithmic scale. \textbf{(Bottom)} Error calculation for optimizing the simulations. The x-axis represents the input angle and the y-axis represents the average error ratio. }
	\label{fig:PSD_different_angles} 
\end{figure}

Figure  \ref{fig:PSD_different_angles} (Top) illustrates the comparison of simulations and measurements for  $\phi_{in} \in \{1^{o},7^{o},10^{o} \}$.  As can be observed, there is a good correlation between the simulations and the measurements. The power given to higher spatial frequencies increases with input angle. This translates into a di\-mi\-nished size of the speckles, i.e. higher spatial fluctuations, which occur from the interference of higher order modes. 
 
A simplified mode coupling procedure with background noise was used to obtain the best correlations between simulations and measurements. This was quantified by optimizing the average error ratio $\epsilon_{dB}$ as illustrated in Fig. \ref{fig:PSD_different_angles} (Bottom). The error of an optimized simulation is compared to a non-optimized procedure. Since the 1D power spectral density plots are expressed in the logarithmic scale, it is useful to express the average ratio in dB with

\begin{equation}
\epsilon_{dB} = 10 \cdot \frac{ \sum_{\nu}^{N}{ \left|\log\left( \mathcal{I}_{mess}(\nu)\right) -  \log\left({\mathcal{I}_{sim}(\nu) } \right)\right|} }{N}.
\label{eq:average_error_ratio_2}
\end{equation}
If $ \rho = |\log\left( \mathcal{I}_{mess}(\nu)\right) -  \log\left({\mathcal{I}_{sim}(\nu) } \right)|$ and $\mu_{\rho}$ represents the mean, it is possible to get the standard deviation in $dB$ through
\begin{equation}
\sigma_{\rho, dB} = 10 \cdot \sqrt{ \frac{1}{N} \sum_{i=1}^{N} \left( \rho_{i}- \mu_{\rho} \right)^{2} }.
\label{eq:standard_deviation_of_error}
\end{equation}
The lower the values $\epsilon_{dB}$ and $\sigma_{\rho,dB}$, the better the correlation between simulations and measurements for the specific input angle.

\subsection{Analysis of angular divergence}
\label{sec:angular_divergence_simulations}

\begin{figure}[htb]
	\centering
	\includegraphics[scale=0.42]{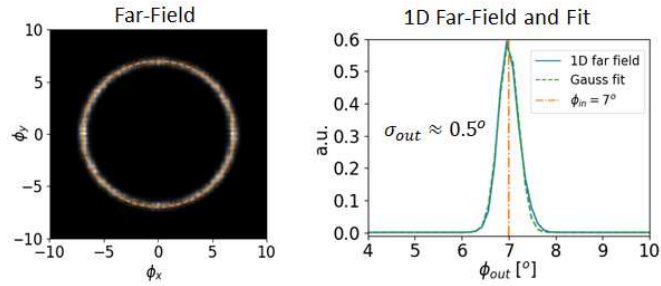}
	\caption{Example of simulated far-field distribution and its correspon\-ding 1D averaged distribution.}
	\label{fig:simulated_far_field_and_1D}
\end{figure}

Simulations of the parallel laser beam method were used to obtain the far-field angular distribution. For the analysis of angular divergence, the resulting 2D far-field intensity distribution is first averaged into a 1D function. Afterwards, the 1D averaged function is fitted by a Gaussian function. To quantify the angular divergence, the standard deviation of the fitted Gaussian function is used as the divergence value. Figure \ref{fig:simulated_far_field_and_1D} shows an example were the 2D far-field distribution is averaged into a 1D distribution and the standard deviation of the Gaussian fit is given. 

\subsubsection{Angular divergence as a function of mode coupling}

\begin{figure}[htb]
	\centering
	\includegraphics[scale=0.43]{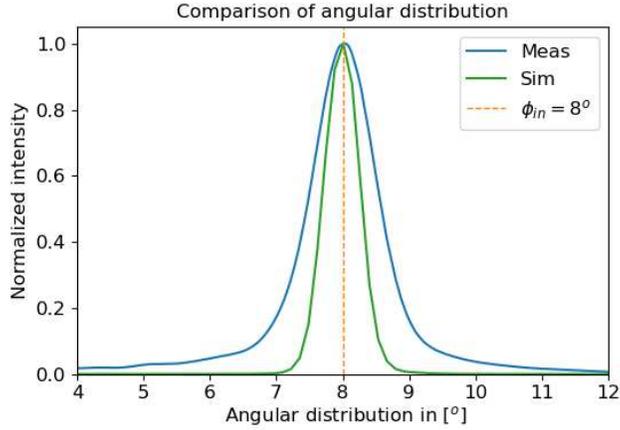}
	\caption{Comparison of the angular distribution between measurement and simulation when using a incident angle of $\phi_{in} = 8^{o}$. The x-axis represents the angular distribution and the y-axis is the normalized intensity.}
	\label{fig:example_1d_mess_sim}
\end{figure}

To realistically simulate the angular divergence, the mode coupling element is needed. For an appropriate value selection of the mode coupling element, a study was conducted using actual measurements of the parallel laser beam method. For each input angle $\phi_{in} \in \{ 6^{o},...,11^{o} \}$, the analysis of the angular divergence was realized, and the results were compared with the simulation. Figure \ref{fig:example_1d_mess_sim} shows measured and simulated results for $\phi_{in} = 8^{o}$.  
  
The real measurement is much more broader than the simulation without mode coupling. Three main phenomena can be attributed to originating FRD: scattering, diffraction and modal diffusion \cite{Haynes11}. The scattering characteristics are best fitted by a Lorentz profile \cite{Haynes11}. In Fig. \ref{fig:example_1d_mess_sim} this can be observed by the long wings of the 1-D profile from the measurement. Modal diffusion and circular aperture diffraction follow a Gaussian distribution \cite{gloge72, Haynes11}. Since the phenomena attributing to the profile in Fig. \ref{fig:example_1d_mess_sim} are described by Gaussian and Lorentz functions, the Voigt profile serves as an appropriate tool.

\begin{figure}[htb]
	\centering
	\includegraphics[scale=0.43]{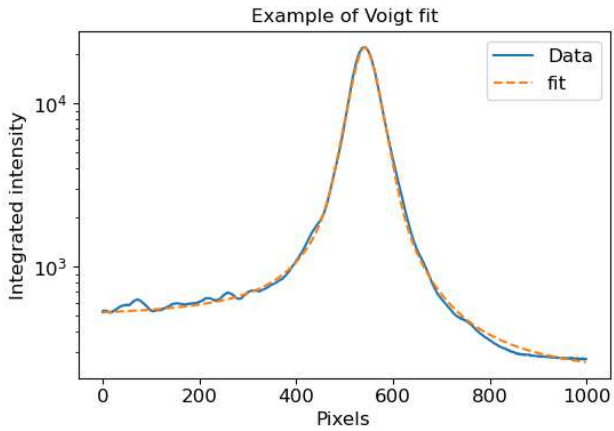}
	\caption{Voigt function applied to the measurement of the circular fibre. The x-axis is in pixels and the y-axis represents the integrated intensity in the log$_{10}$-scale.}
	\label{fig:voig_ciruclar_fixed}
\end{figure}

For ease of calculations, a pseudo-Voigt function $\mathcal{V}(r)$ was used, which represents a linear approximation that provides an analytical integrable combination of Gaussian and Lorentzian functions. The validity of the pseudo-Voigt function has been proven by many authors \cite{Whiting68,Peyre72,Kielkopf73,Ida00}. The approximation used for this study is the one described in \cite{Ida00}. Figure \ref{fig:voig_ciruclar_fixed} shows the result of the fit using an actual measurement.  

\begin{figure}[htb]
	\centering
	\includegraphics[scale=0.45]{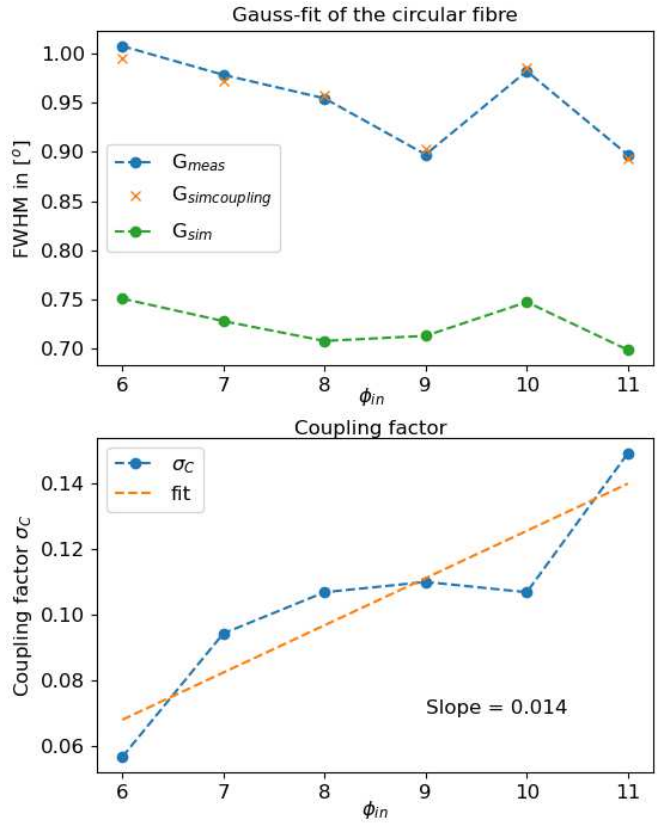}
	\caption{\textbf{(Top)} Results of angular divergence between simulations and measurements for $\phi_{in} \in \{6^{o},...,11^{o}\}$. The x-axis represents the input angle $\phi_{in}$, and the y-axis is the angular divergence illustrated as the FWHM of the Gaussian fit. \textbf{(Bottom)} Mode coupling factor $\sigma_{C}$ needed to match $G_{meas}$.}
	\label{fig:Coupling_factor_FWHM_sim_meas}
\end{figure}

The FWHM of the Gaussian component can be calculated with $\Gamma_{G} = 2\sigma_{G}\sqrt{\text{ln}2}$, with $\sigma_{G}$ being the standard deviation of the Gaussian fit function. To separate the modal diffusion component, it is possible to apply the addition of two Gaussians with $\sigma_{D} = \sqrt{\sigma_{G}^{2} - \sigma_{A}^{2}}$. The circular aperture diffraction parameter is represented with $\sigma_{A}$ and is the same for both the measurements and the simulations. Circular aperture diffraction can be described with the Airy disk. The angle subtended by the radius of the Airy disk is $\theta_{Airy} = 1.22\frac{\lambda}{D}$, where $D$ is the diameter of the aperture.
Since the central part of the Airy disk can be approximated by a Gaussian, the standard deviation of the Airy disk is $\sigma_{A} = \frac{1}{2}\theta_{Airy}$ 
The modal diffusion component can be simulated using the simplified mode coupling procedure. The results of the angular divergence are given as the FWHM of the modal diffusion component. The input field diameter for the simulations was $d_{in} = 50\mu m$, matching the diameter of the fibre. The mode coupling component was incremented until $\sigma_{G,sim} = \sigma_{G,mess}$. Figure \ref{fig:Coupling_factor_FWHM_sim_meas} shows the results: $G_{sim}$  are simulations without mode coupling, $G_{simcoupling}$ the simulations with mode coupling, and $G_{meas}$ the measurements. The root mean square error (RMSE) was calculated as 
\begin{equation}
RMSE = \sqrt{\frac{1}{n}\sum^{n}_{ii}(G_{meas,ii} - G_{simcoupling,ii})^{2}}
\label{eq:RMSE}
\end{equation}
and equals $0.0064^{o}$.

\begin{figure}[htb]
	\centering
	\includegraphics[scale=0.45]{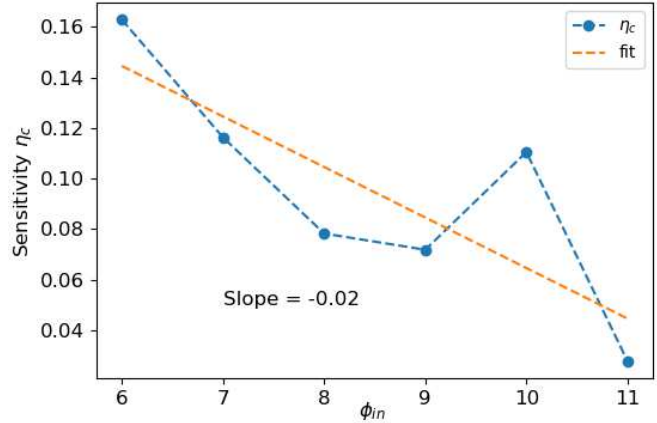} 
	\caption{Sensitivity of the input angle with respect to the coupling component. The x-axis shows the input angle and the y-axis the sensitivity $\eta_{c}$. A higher value indicates higher sensitivity.}
	\label{fig:sensitivity}
\end{figure}

It is possible to match the angular divergence quite well using the simplified mode coupling component. There is also a tendency for the coupling to increment with the incident angle. To quantify the sensitivity $\eta_{c}$ of the individual input angles to the coupling factor $\sigma_{c}$,  the angular divergence $G_{sim,\sigma}$ for the each input angel was simulated using the range of values for $\sigma_{c}$ that can be observed in Fig. \ref{fig:Coupling_factor_FWHM_sim_meas}. The final result for $\eta_{c}$ is given as the standard deviation of the resulting $G_{sim,\sigma}$ pro input angle. Figure \ref{fig:sensitivity} shows the results.  

A higher value of $\eta_{c}$ indicates more sensitivity to the coupling factor. This means that the angular divergence experiences stronger fluctuations with changes in the coupling factor $\sigma_{c}$. Since the coupling factor can be accounted for all the disturbances in the systems that produce mode coupling,  these results indicate that lower input angles are more sensitive to disturbances. 

\subsubsection{Angular divergence as a function of the input field diameter}

The influence of the input field diameter on the resulting far-field angular distribution was examined by simulations. As fluctuations in the input spot diameter translate into divergence in the angular distribution, which in turn cause FRD, fluctuations in the input spot diameter have to be avoided. However, this situation typically occurs in reality due to seeing variations, defocus, or poor guiding of the telescope, i.e. misalignment of the fibre in the focal plane.

\begin{figure}[tbh]
	\centering
	\includegraphics[scale=0.38]{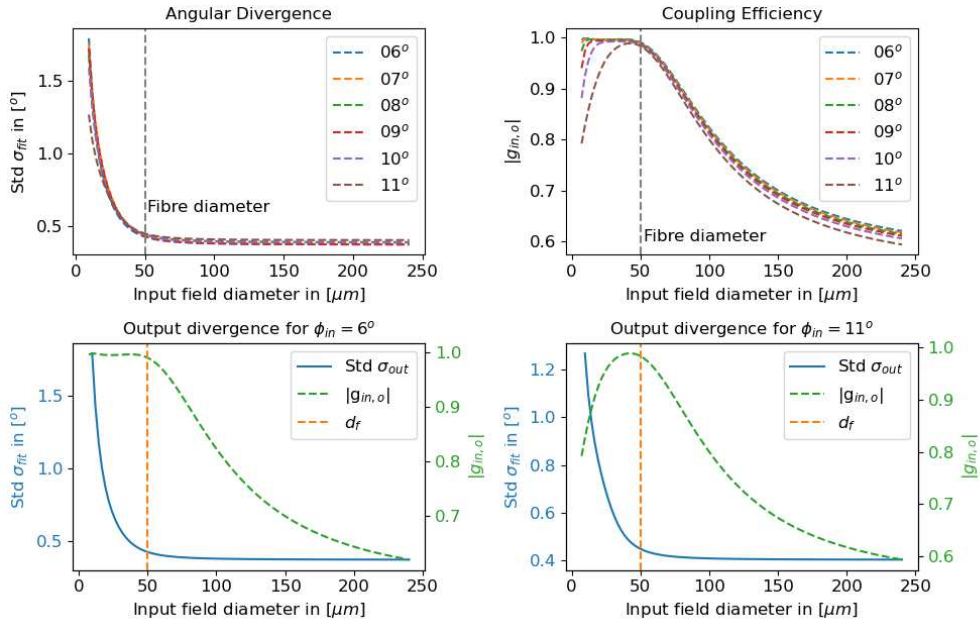}
	\caption{\textbf{(Top-left)} Angular Divergence: The y-axis represents the angular divergence and the x-axis the increasing input field diameter. \textbf{(Top-right)} Coupling Efficiency: The y-axis represents the coupling efficiency and the x-axis the increasing input field diameter. \textbf{(Bottom)} Output Divergence: Individual examples with $\phi_{in} = 6^{o}$ and $\phi_{in} = 11^{o}$.}
	\label{fig:output_divergence_v02} 
\end{figure}

To run a simulation, the input angle $\phi_{in}$ is set to a specific value and afterwards the input field diameter is increased from 0.1$d_{f}$ to $\sim 5d_{f}$ in n = 100 steps, with $d_{f}$ indicating the fibre diameter. The resulting far-field annular distribution is transformed to a 1-D function and is fitted to a Gaussian function. The angular divergence is calculated for each step and is given as the standard deviation of the Gaussian fit $\sigma_{f\!\!it}$, as already illus\-trated in Fig. \ref{fig:simulated_far_field_and_1D}. The top left part of Fig. \ref{fig:output_divergence_v02} shows the results for the input angles $\phi_{in} \in \{ 6^{o},...,11^{o}\}$. There is a minimal difference between the individual calculations, but the same trend is followed. The angular divergence increments rapidly after the field diameter is smaller than the fibre diameter. For field diameters bigger than the optical fibre, a limit is reached quite rapidly for all input angles. This describes that underfilled fibres will be more susceptible to variations of the input field diameter. The coupling efficiency as a function of the input field dia\-meter was also analyzed. The spatial correlation between the input field and the field produced at $\Psi(x,y, z = 0)$ is  $G_{1,2} = \langle \psi_{1}^{*}, \psi_{2} \rangle$
and is normalized to 
\begin{equation}
g_{1,2} = \frac{G_{1,2}}{\sqrt{I_{1} I_{2}}}.
\label{eq:normalized_mutual_intensity}
\end{equation}
The absolute value of $g_{1,2}$ is bounded between zero and unity, $0 \leq |g_{1,2}| \leq 1$. If $|g_{1,2}|$ = $1$, both fields are identical  $(\psi_{1} = \psi_{2})$. As described in Eq. \ref{eq:input_z_equals_zero}, the input field  can be calculated for $z=0$ as,
\begin{equation}
\Psi_{in} = \sum_{m=0}^{M} c_{m}\psi_{m} = \psi_{o}.
\label{eq:psi_o_mutual}
\end{equation} 
Since only a finite number of propagation modes are available, Equation \ref{eq:psi_o_mutual} can be described best with $\Psi_{in} \approx \psi_{o}$. If the normalized mutual intensity is applied to both $\Psi_{in}$ and $\psi_{o}$, then $|g_{in,o}|$ gives a reliable statement in the similarity between both fields. The top right part of Fig. \ref{fig:output_divergence_v02} shows the results for the input angles $\phi_{in} \in \{ 6^{o},...,11^{o}\}$. For all input angles, the coupling efficiency decreases when the field diameter surpasses the fibre diameter. For larger input angles, it appears that the coupling efficiency is also affected by decreasing the field diameter, which leads to higher sensitivity to fluctuations in underfilled fibres.

With this information it is possible to produce plots for analyzing the angular divergence and coupling efficiency of individual input angles for a given system. This is illustrated in the bottom part of Figure \ref{fig:output_divergence_v02} for the input angles $\phi_{in}\in {6^{o},11^{o}}$.


\section{Application}
\label{sec:application}
At the early design stages of a fibre-coupled spectrograph system, it is important to understand the fibre properties as they have an important impact on the overall efficiency and stability of the system. While traditionally efficiency has been estimated in the past with figures of merit, e.g.\ FRD data, allowing to distinguish between more or less performing fibre types, the temporal stability with regard to modal noise and scrambling is more of a challenge, and often left to tweaking an integrated system after completion with trial and error for optimal performance. Our mode expansion theory based model allows to simulate long fibres, and study the behaviour depending on different conditions of light injection into the fibre, as well as connection to the spectrograph optical system. Although an optical waveguide can propagate only a finite number of modes, a realistic fiber with a typical length of, say, 30m is far from an idealized cylindrical waveguide, that, in principle, could be subject of a fully deterministic description. However, our model, coupled to a ray tracing model of the optical system of the spectrograph, and resulting in a full end-to-end model, allows parameter studies that are valuable for system design and optimization. 

A good example for such an approach can be found in the detailed radial velocity error budget for the NASA-NSF Extreme Precision Doppler Spectrometer instrument concept NEID \cite{Halverson2016}. This study finds that imperfect scrambling of the near-field results in uncalibrated velocity shifts in the spectrometer focal plane with adverse effects on the accuracy of radial velocity measurements. Imperfect scrambling in the far-field results in variations of the spectrometer pupil illumination, redistributing the optical system aberrations that in turn are the cause for systematic PSF shifts in the focal plane. These considerations are exactly the motivation for our end-to-end simulation approach.

\begin{figure}[t]
	\centering
	\includegraphics[scale=0.32]{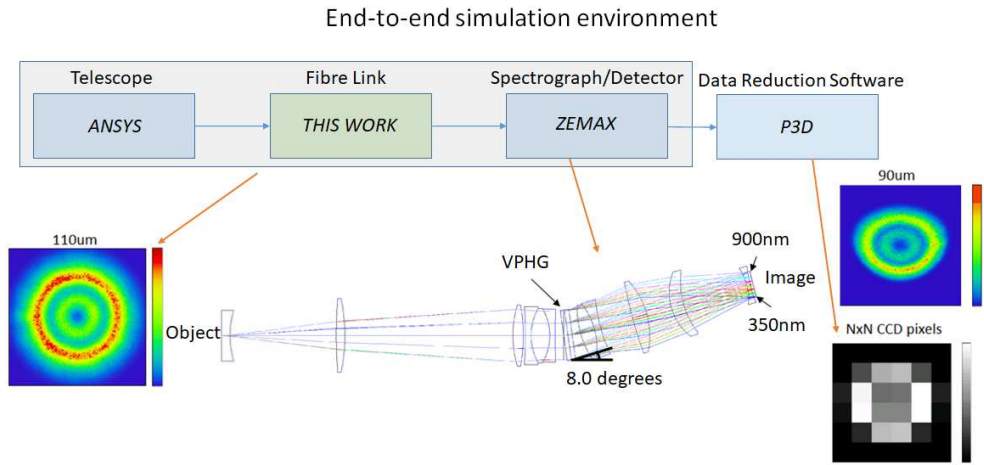}
	\caption{\textbf{(Top)} Block diagram  of end-to-end simulation.  \textbf{(Bottom)} Illustration of an end-to-end simulation. An intensity distribution is used at the input end of the spectrograph. The output end shows the resulting distribution after ray tracing through Zemax is applied. The resulting 2D image is sampled on the detector.}
	\label{fig:end_end_simulations} 
\end{figure}

The end-to-end simulation environment is realized by coupling the output of the simulated near- and far-field light distributions to a Zemax model of the spectrograph. This includes proper treatment of the fibre core illumination at the location of the pseudo-slit, as well as the illumination of the spectrograph pupil, determined by the far-field. This task is accomplished by special DLL routines developed for the Zemax code.

Figure \ref{fig:end_end_simulations} shows a demonstration of this process with a fibre-fed version of a MUSE-like spectrograph design obtained from  \cite{Moralejo16}. A simulated intensity distribution can be seen at the object plane of the spectrograph. The field is propagated and the resulting distribution is delivered at the image plane. The image is sampled by the CCD detector. Subsequent data analysis is performed with the multi-fibre spectroscopy data reduction software package P3D\footnote{https://p3d.sourceforge.io/} \cite{Sandin10}. By recording hundreds of such data points over different field angles along the slit, dispersed by the spectrograph, we can sample an entire fibre-coupled IFU. It is then possible to immediately study the impact of modal effects, e.g.\ caused by different ways to illuminate fibers at in telescope focal plane, on the LSF of the resulting spectra, and assess system properties like the accuracy of Doppler shift measurements, sky emission line subtraction accuracy, etc.

This kind of end-to-end simulations is not limited to the above example. Adaption to any kind of optical system is possible, thus opening a wide field of applications. A description of the Zemax DLLs and results from a comprehensive study will be reported in a forthcoming paper.

\section{Summary and conclusions}
\label{sec:summary}
This works simulates the light propagation in step-index multimode fibres and the multimode effects with a rigo\-rous treatment of wave propagation. Using the modes as a basis for linear expansion and taking advantage of the harmonic z-dependence of the propagation modes, it was shown that it is possible to calculate the propagation in multimode fibres in an effective and efficient way. 
It  was also shown that it is possible to analyze the output fields of fibres by inserting spatial fluctuations, angular fluctuations, focus varia\-tions, with different kinds of fibre cross-sections, propa\-gation wavelengths, diameters and numerical apertures.

Insights into the intrinsic causes of the complex effects that diminish the efficiency in fibre-based system were given. The resulting increase of angular divergence as a function of modal redistribution was presented by the different input parameters and using the simplified mode coupling procedure. This is analogous to an increase in FRD that a system will experience.
As was demonstrated, the output intensities of both the near-and far-field vary due to many input parameters. This indicates and reaffirms the difficulties, e.g. for a precise fibre-based sky background subtraction method, since every fibre will have a slightly different modal power distribution. Perfect modal scrambling in fibres could ease the differences in the modal power distribution. 
The effects of modal noise as a function of the input angle were demonstrated accurately in the near-field. 

The predictive power of the model is limited to the fact that the calibration of the mode coupling parameter is realized by using actual measurements. An advantage that arises, is the possibility of examining the obtained modal power distribution. Afterwards, perturbances can be applied to the modal power distribution to see how much the output intensity fields may vary. This gives a predictive power into the range of changes that might be expected in both the near- and far-field for a specific fibre-fed system or setting. Although  there is still work to be done, the simulation environment can already be used as a  tool to assist with the characterization and design process for fibre-fed spectrograph systems.


\begin{backmatter}
\bmsection{Funding}
This work was funded by \emph{BMBF} Grant 05A14BA.

\bmsection{Acknowledgments}
This work was supported the research and innovation centre innoFSPEC. 

\bmsection{Disclosures}
The authors declare no conflicts of interest.

\bmsection{Data Availability Statement (DAS)}
Data underlying the results presented in this paper are not publicly available at this time but may be obtained from the authors upon reasonable request.

\bmsection{Supplemental document}
See supplemental document for supporting content describing the experimental setup. 

\end{backmatter}


\bibliography{Bibliography}

\end{document}